\documentclass[sigconf]{acmart}

\usepackage{graphicx}
\usepackage{subcaption} 
\AtBeginDocument{%
  }

\setcopyright{acmlicensed}
\copyrightyear{2026}
\acmYear{2026}
\acmDOI{10.1145/3794786.3830772}
\acmConference[GoodIT ’26]{International Conference on Information Technology for Social Good}{September 2--4,
  2026}{Pisa, Italy}
\acmISBN{978-1-4503-XXXX-X/2026/06}

\usepackage{booktabs}  
\usepackage{multirow}

\usepackage{tikz}
\usetikzlibrary{arrows.meta,positioning}

\tikzset{
  box/.style={
    rectangle,
    rounded corners=2pt,
    draw,
    thick,
    align=center,
    inner sep=6pt,
    font=\footnotesize
  },
  rootbox/.style={box, draw=green!60!black, fill=green!10},
  issuesbox/.style={box, draw=blue!70!black, fill=blue!5},
  ideasbox/.style={box, draw=magenta!70!black, fill=magenta!5},
  addressbox/.style={box, draw=orange!80!black, fill=orange!10},
  teachbox/.style={box, draw=olive!80!black, fill=olive!10},
  subblue/.style={box, draw=blue!70!black, fill=blue!5},
  subpink/.style={box, draw=magenta!70!black, fill=magenta!5},
  arrow/.style={-Latex, very thick, draw=gray!70}
}

\begin{document}

\title[Making Gender-Inclusive Practices Actionable]{Making Gender-Inclusive Practices Actionable: Evaluating a Research-Informed Computing Education Toolkit}

\author{Alina Berry}
\email{alina.berry@tudublin.ie}
\orcid{0000-0003-1528-7676}
\affiliation{%
  \institution{Technological University Dublin}
  \city{Dublin}
  \country{Ireland}
}

\author{Susan McKeever}
\affiliation{%
  \institution{Technological University Dublin}
  \city{Dublin}
  \country{Ireland}}

\author{Brenda Murphy}
\affiliation{%
  \institution{SETU Waterford}
  \city{Waterford}
  \country{Ireland}
}

\author{Sarah Jane Delany}
\affiliation{%
 \institution{Technological University Dublin}
 \city{Dublin}
 \country{Ireland}}

\renewcommand{\shortauthors}{Berry et al.}

\begin{abstract}

The persistent gender imbalance in computing remains a global concern, and universities offer a key part of the pipeline to address it. Although research has identified practices that support under-represented student groups, translating this evidence into actionable guidance remains challenging. This paper first presents a novel web‑based toolkit (TechMate) designed to address this gap by helping computing educators implement gender‑inclusive initiatives through practical research-informed guidance. The toolkit defines over 25 actions, ranging from operational to strategic, and provides case studies and implementation resources.
Second, this work reports on the evaluation of TechMate, capturing educators' first impressions of its usefulness and usability through authentic tasks, and eliciting unanticipated insights about structural barriers to gender-inclusive practice in computing higher education. Eighteen computing educators of varying seniority from eight universities in Ireland assessed the toolkit using nine research‑driven attributes: four usefulness measures (\textit{novelty}, \textit{relevance}, \textit{trustworthiness}, \textit{actionability}) and five usability measures (\textit{effectiveness}, \textit{aesthetics}, \textit{navigation}, \textit{terminology}, \textit{user satisfaction}). Data was analysed using mixed methods: deductive coding, quantitative sentiment scoring, and inductive thematic analysis.
Educators rated TechMate highly for \textit{user satisfaction} and praised its \textit{novelty} and \textit{trustworthiness}, with over 60\% of participants reporting positively on all attributes, suggesting that research‑informed tools can be designed for real‑world use. Thematic analysis highlighted other challenges, including women students’ isolation, institutional resistance to change, and the scarcity of women role models, while also surfacing proposed solutions such as leveraging student ambassadors and fostering supportive learning environments.
This work contributes to the broader question of how open-access digital tools can support equitable participation in computing, with implications for design and governance in education and social good contexts.

\end{abstract}





\maketitle

\section{Introduction}

The persistent under-representation of women in computing is well documented: women\footnote{To include relevant gender minorities (such as, for instance, trans women), the collective term ``women'' is used in this paper to describe the gender identity, as opposed to using the term ``female'' that would only include those who were born as women.} make up less than 20\% of ICT (Information and Communication Technologies), specialists across the EU (European Union) \cite{EuroStat1}, and almost half leave technical careers before age 35, reflecting the "leaky pipeline'' \cite{Accenture2020ResettingTech}. This pattern is visible in higher education; for example, only 21\% of students in introductory computing courses in Ireland and Denmark were women \cite{Quilleetal2017}. 

Gender imbalances in computing deprive the field of diverse talent and the associated advantages of heterogeneous teams, such as improved innovation, creative problem solving, and collective intelligence \cite{Page2007, Woolley2010}. In Ireland, which has Europe's largest gender gap in advanced digital skills use at work \cite{ESRI2026}, addressing the issue is essential not only for gender equality but also for productivity and inclusive economic growth.

The under-representation of women in computing emerges from multiple factors, including early educational experiences that shape stereotypes, self-efficacy, values and interests \cite{Beyer2014}, broader societal and cultural influences \cite{cheryan2017some}, and structural barriers embedded in educational and workplace environments \cite{Margolis2002}. Although these challenges originate well before university and persist beyond graduation, computing educators nonetheless have responsibility and opportunity to implement research-informed practices that enhance recruitment, support retention, and foster inclusive experiences for women and other under-represented groups.

This paper presents and evaluates TechMate, a freely available web-based toolkit designed to help computing educators implement gender-inclusive initiatives through practical, research-informed guidance. Unlike high-level frameworks or policy documents \cite{EIGE, OECDtoolkit}, TechMate emphasises practicality, offering ready-to-use resources that educators can access on demand. Beyond its immediate contribution to computing education, this work speaks to a wider question relevant to technology for social good: under what conditions can open-access digital tools actually shift institutional practice, rather than simply documenting what good practice looks like?

The focus of this paper is to establish TechMate’s design and first‑use value rather than its long‑term impact on gender ratios or student outcomes, with those effectiveness studies identified as future work building on this initial stage.

The evaluation of TechMate addresses three research questions: \textbf{(RQ1)} \textit{How do educators perceive the usefulness of the toolkit's content?} \textbf{(RQ2)} \textit{How usable is the toolkit for computing educators?} and \textbf{(RQ3)} \textit{What unanticipated needs or insights emerge from users' interactions with the toolkit?}

This paper is structured as follows. Section \ref{background} reviews related work; Section \ref{toolkit} introduces the toolkit; Section \ref{methodology} describes the evaluation study design; Section \ref{results} reports findings and implications; Section \ref{conclusion} concludes with limitations and directions for future work.

\section{Background/Related Work} \label{background}

Gender imbalance in computing stems from multiple interacting factors. Beyer's work \cite{Beyer2014} identifies stereotypes, self-efficacy beliefs, values, and interests as key predictors of women's participation, demonstrating that these psychological factors are shaped by educational and social experiences from early schooling onwards. Cheryan et al. \cite{cheryan2017some} provide evidence that computing's masculine culture and narrow stereotypes deter women more than cognitive ability differences, while Margolis and Fisher's foundational ethnographic study \cite{Margolis2002} documents how institutional structures, classroom dynamics, and cultural assumptions systematically disadvantage women in computer science education.

It is also important to acknowledge critical perspectives on this body of work. Vitores and Gil-Juárez \cite{vitores2016} argue that ``women in computing'' as a research framing carries risks: by focusing on why women fail to enter or persist in computing, it can reproduce a deficit model that locates the problem in women themselves rather than in the institutional and structural conditions that shape their exclusion. Perez-Felkner et al.'s \cite{PerezFelkner2024} systematic review of computing interventions from 2000 to 2020 reinforces this, finding that intervention efficacy depends on social and structural factors as much as on academic ones. 

While substantial research documents effective interventions to support under-represented gender groups in computing education, including strategic recruitment initiatives, as for example, women only programmes \cite{siegeris9mehr} and rebranding of the introductory computing course \cite{wallace2016girls}, curriculum redesign with hybrid programmes \cite{brodley2022, mckeever2021addressing}, and mentoring schemes \cite{dennehy2017female, Sigplan}, this evidence remains dispersed across academic and grey literature \cite{minerva}. Although several reviews synthesise intervention strategies \cite{goos2020review, Morrison2021, happe2021effective}, translating research findings into accessible, actionable guidance for educators continues to be a challenge. 
Henderson et al. \cite{Henderson2011} identify this research-practice gap as a persistent barrier to instructional change in undergraduate STEM education, noting that even when evidence-based practices are documented, structural and cultural factors impede their adoption, a phenomenon observed across other educational domains \cite{Graham2006}.

In response, various toolkits and frameworks have been developed to support educators. NCWIT's Undergraduate System Model and Engagement Practices Framework \cite{NCWITUndergraduate, NCWIT2025} provide conceptual guidance for diagnosing structural factors and compiling research-based practices, though they function as separate resources requiring interpretation across multiple documents. Taylor-Smith et al.'s Participant-Centred Planning Framework \cite{taylor-smithetal2022} supports activity design and evaluation but focuses primarily on use of role models and does not provide the comprehensive collection of ready-to-implement interventions with concrete resources that educators need to translate research into practice. International organisations have developed broader gender equality toolkits \cite{EIGE, OECDtoolkit}, discipline-specific guides \cite{moredapozo2020csframeworks, spieler2025pecc}, and institutional toolboxes \cite{toolboxNtnu, GENDIMNorway}. While valuable, these resources present limitations: many provide high-level strategic frameworks without detailed implementation specifics, some target specific institutional contexts limiting transferability, and others lack direct connections to supporting research evidence or provide limited guidance for institutional implementation.
Additionally, any systematic evaluation of their usability and educator experience remains limited, with studies focusing on intervention effectiveness \cite{achenbach2018systemic, thompson2021increasing} rather than the digital resources that curate and present them.
Although established models such as the Technology Acceptance Model \cite{Davis1989}, Nielsen's heuristics \cite{Nielsen1993}, and the System Usability Scale \cite{Brooke1996} are widely used to assess digital systems, their application to gender-focused educational toolkits is scarce. As a result, little is known about how educators engage with such resources in local contexts or which design features might support the translation of evidence into practice.

\section{The Toolkit - TechMate}\label{toolkit}

TechMate \cite{TechMate} was designed to address the identified gap between research evidence and educator practice directly. Unlike workshops, which are temporal and require attendance, or policy documents, which are typically high-level and disconnected from day-to-day practice, a web-based toolkit provides persistent, on-demand access to actionable resources that educators can consult at their own pace and return to as their institutional context evolves.

Additionally, it is worth contrasting TechMate with some of the other practical resources (as mentioned in Section \ref{background}) designed to address the same issue. Unlike NCWIT’s Engagement Practices Framework, which appears to focus mainly on in-class initiatives \cite{NCWIT2025}, TechMate covers a range of levels for participation, from educators to policy makers and industry professionals. The NCWIT's Undergraduate System Model \cite{NCWITUndergraduate} builds on six focus areas: Data \& Evaluation, Program Entry, Classroom Experience, Program Curriculum \& Supports, Community \& Belonging, and Institutionalisation, and offers some concrete practical ideas in each. However, it is tailored specifically to undergraduate students in the US and does not consistently link research evidence directly to individual recommendation pages. In addition, users must navigate multiple pages and external, more generic resources, so the material functions more as a dispersed set of guidance documents than as a single, unified toolkit. TechMate has been developed in Irish context but it also makes a case for using experience from other countries, allows for a postgraduate cohort as well as undergraduate and prospective students, and presents each initiative on a single, self‑contained page that combines all resources.

\begin{figure}[h]
  \centering
  \includegraphics[width=1\linewidth]{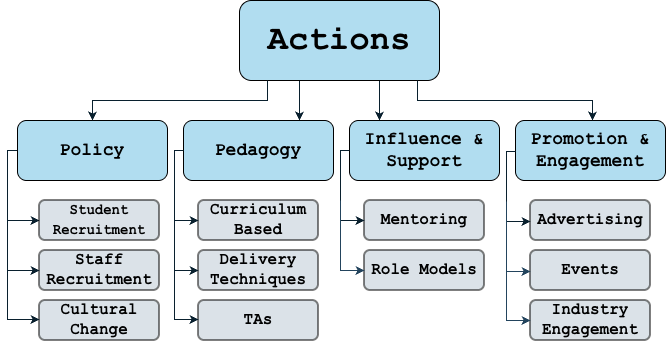}
  \caption{Four categories of actions in TechMate with respective subcategories.}
  \label{fig:categorisation_anonymised}
\end{figure}

It organises 25+ initiatives, called actions (with new actions added as research evolves) into four high-level categories (derived from prior work \cite{BERRY2022ADD}) reflecting \textit{Policy} changes, \textit{Pedagogical} practices, \textit{Influence \& Support} mechanisms, and \textit{Promotion \& Engagement} approaches to addressing gender imbalance in computing higher education. 
Figure~\ref{fig:categorisation_anonymised} shows the four categories with respective subcategories. Examples include: gender-balanced staff recruitment and marketing redesign (\textit{Policy}); hybrid programmes and personalised feedback (\textit{Pedagogy}); buddy systems and peer mentoring (\textit{Influence \& Support}); and school outreach events and industry partnerships (\textit{Promotion \& Engagement}).

\begin{figure}[h]
  \centering
  \includegraphics[width=1\linewidth]{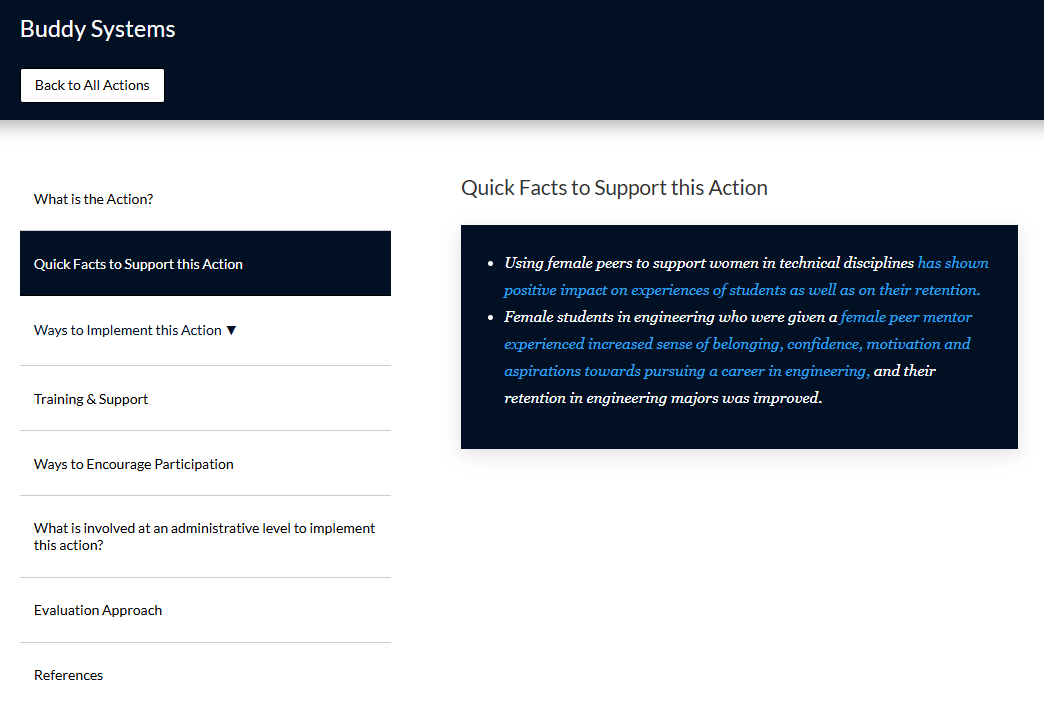}
  \caption{Buddy Systems page in TechMate (screenshot).}
  \label{fig:screenshot_TechMate}
\end{figure}

Each intervention is presented as a discrete action, accompanied by concise supporting evidence, practical implementation guidance, and suggested evaluation approaches to encourage reflective and data-informed adoption.
Each action page follows a consistent structure with six core sections: a description of the action, quick facts summarising research evidence, ways to implement with concrete steps, evaluation approaches with sample instruments, ready-to-use resources such as templates, and references to relevant sources.
For example, the action \textit{Buddy Systems} (\textit{Mentoring} subcategory within the \textit{Influence \& Support} category) includes techniques on how to pair junior women students with more senior women peers to support their transition into a men-dominated computing environment. The action page opens with a concise description and research evidence: how women peer mentors have been shown to increase women's sense of belonging, confidence, and retention in technical disciplines \cite{dennehy2017female}. The implementation section then walks through recruitment (advertising the scheme before the academic year begins, contacting all newly enrolled women students, and sourcing buddies through faculty recommendations), matching strategies (by shared interests, randomly, or by preference), training (a two to three hour session structured around buddies sharing their own first-year experiences), and participation incentives (stipends, awards, or informal rewards). An evaluation approach is suggested, recommending pre- and post-scheme surveys for mentees and reflective journals for buddies after each meeting. Ready-to-use resources, including example training handouts and application forms drawn from institutions such as the University of York and Carleton University, are linked directly from the page. Figure \ref{fig:screenshot_TechMate} shows a screenshot of the Quick Facts view.

The action \textit{Class/Lab Dynamics} (\textit{Delivery Techniques} subcategory, \textit{Pedagogy} category) focuses on managing in‑class interactions so that women and other minoritised students are not left out during practical work. It suggests concrete steps such as, for example, placing a woman student to work with at least one other woman, addressing students by their names, and using brief check‑ins to ensure that quieter students have opportunities to contribute, all supported by research on participation patterns in computing classrooms. The action page also contains survey suggestions aimed to evaluate impact.

\begin{figure*}[t]
  \centering
  \includegraphics[width=0.85\textwidth]{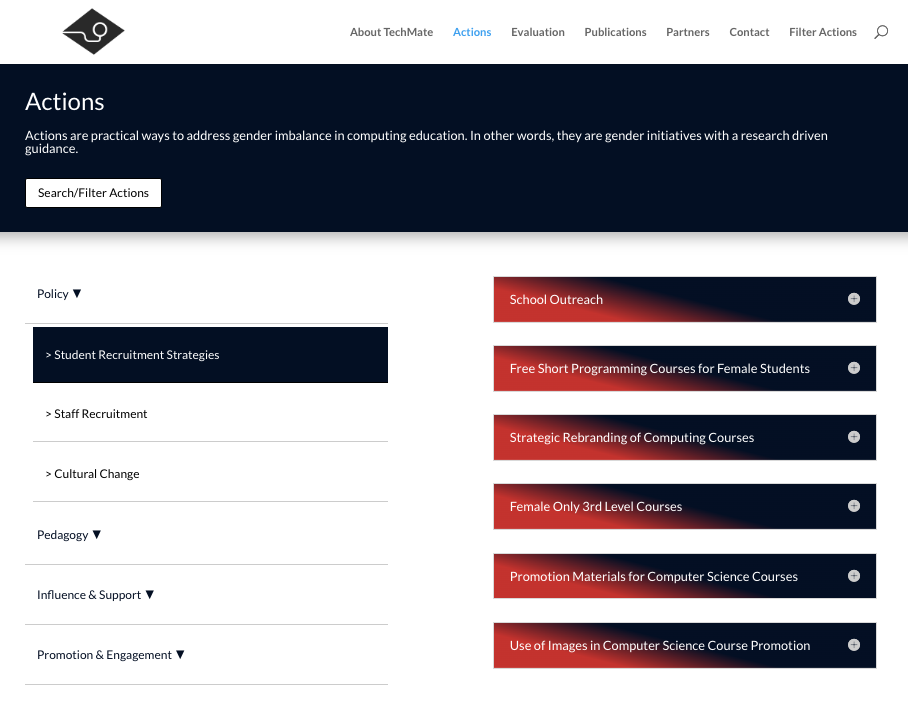}
  \caption{Screenshot from TechMate displaying navigation through categories and sample actions for one of the subcategories.}
  \Description{Screenshot from TechMate displaying navigation through categories and sample actions for one of the subcategories.}
  \label{fig:categories_sample_actions}
\end{figure*}

TechMate provides two discovery pathways for users: category-based navigation through an interactive side menu displaying all four categories and their subcategories (as demonstrated in Figure \ref{fig:categories_sample_actions}), suitable for first-time visitors seeking comprehensive understanding; and filter-based search enabling targeted discovery based on the user's \textit{ultimate goal} (recruitment or retention), the \textit{level of impact} (short-term or long-term), the \textit{student cohort targeted} (new, undergraduate, or postgraduate students), and the \textit{role of the user} (lecturer, manager, policy maker, or other). This dual approach accommodates different user preferences, expertise levels, and institutional contexts.

\section{Evaluation Methodology}\label{methodology}

The aim of the evaluation of TechMate was to assess first impressions on its perceived usefulness (RQ1) and usability (RQ2) among computing educators through theory-driven attributes and elicit any unexpected insights (RQ3) during evaluation sessions.

\textbf{Perceived usefulness}, grounded in the Technology Acceptance Model, is defined as both the degree to which someone believes that an application will help them do their job better \cite{Davis1989} and as the usefulness of the content \cite{kushniruk2004cognitive}. Although the primary job of the target audience of the toolkit does not typically include implementing gender initiatives, the toolkit’s aim is to help those interested in or focused on achieving better gender balance, making it easier to do this. To be perceived as useful, the toolkit should offer a degree of novelty and new ideas on how to enhance gender balance in computing higher education. The information should also be relevant to users with different levels of decision making authority. In addition, it should provide reliable and trustworthy sources supported by academic research. Finally, the content of the toolkit should be actionable, so those seeking help can find everything required to start implementing its guidance. Hence, the \textbf{usefulness} assessed in this context comprised four attributes: \textit{novelty}  \cite{Davis1989, Rogers2003} (new and useful ideas), \textit{relevance} \cite{buchanan2009evaluating, Barry1998, Tombros2005, Wang1996} (appropriateness to user roles), \textit{trustworthiness} \cite{Wathen2002, Tsakonas2006} (source credibility), and \textit{actionability} \cite{Wang1996, Venkatesh2003} (clarity and feasibility of proposed guidance).

The usability standard ISO 9241-11 \cite{ISO9241-11} defines \textbf{usability} through three key attributes: effectiveness, efficiency, and satisfaction. There are studies that use additional attributes to test usability, such as memorability, learnability, errors \cite{nielsen2012usability}, or aesthetic appearance, navigation, and terminology \cite{buchanan2009evaluating}. Usability evaluation is commonly conducted with the help of user tasks that are performed in the assessed application \cite{ISO9241-11, Nielsen1993}, ensuring that the interface and information presentation provide an adequate cognitive fit for user tasks \cite{Vessey1991}.
For this study, \textbf{usability} was captured using five attributes: \textit{effectiveness} \cite{ISO9241-11, Brooke1996, buchanan2009evaluating} (ease of completion of evaluation tasks), \textit{aesthetics} \cite{Hassenzahl2003, Nielsen1994} (visual design quality), \textit{navigation} \cite{Barnes2002, Nielsen1994} (movement between pages), \textit{terminology}  \cite{Nielsen1994, Sweller1988} (language clarity) and \textit{user satisfaction} \cite{Brooke1996, Chin1988} (overall experience).

Ethical approval was obtained from Technological University Dublin prior to data collection.

\begin{table*}[t]
\centering
\caption{Individual scores for each attribute for coders S1, S2 and S3, the average score and illustrative participant quotes for three participants, P01, P06 and P14.}
\label{tab:attribute-scores-quotes-reduced}
\footnotesize
\begin{tabular}{p{2.0cm}p{0.6cm}ccccp{9.0cm}}
\toprule
\textbf{Attribute} & \textbf{Part.} & \textbf{S1} & \textbf{S2} & \textbf{S3} & \textbf{Avg.} & \textbf{Representative Meaningful Unit} \\
\midrule
\multirow{2}{*}{Novelty} & P06 & 4 & 4 & 4 & 4 & \textit{"So there's some good ideas there about how you would evaluate that."} \\ & P14 & 4 & 3 & 4 & 4 & \textit{"Research shows computing courses worded with people‑oriented keywords attract more women applicants."} \\
\midrule
\multirow{2}{*}{Relevance} & P06 & 5 & 4 & 5 & 5 & \textit{"Yeah, very much so. Absolutely. Very applicable."} \\ & P01 & 5 & 4 & 5 & 5 & \textit{"What you have here is exactly what we… care about for a larger, diverse population in computer science."} \\
\midrule
\multirow{2}{*}{Trustworthiness} & P06 & 5 & 4 & 5 & 5 & \textit{"I did trust what I read because I trust you… I would have good faith in the information on here."} \\ & P14 & 5 & 4 & 5 & 5 & \textit{"It'd be interesting to know what that research is… there's the link, excellent."} \\
\midrule
\multirow{2}{*}{Actionability} & P06 & 5 & 4 & 4 & 4 & \textit{"You've got information about how you could implement and evaluate the actions you've taken."} \\ & P01 & 5 & 4 & 5 & 5 & \textit{"I would spend time reading this and taking the action items for what I can do."} \\
\midrule
\multirow{2}{*}{Effectiveness} & P06 & 5 & 5 & 5 & 5 & \textit{"[Tasks] easy, very easy to follow."} \\ & P14 & 5 & 4 & 3 & 4 & \textit{"That was very straightforward. I had no problem with the task."} \\
\midrule
\multirow{2}{*}{Aesthetics} & P06 & 5 & 4 & 5 & 5 & \textit{"Not too much clutter. Nicely designed, very clear and easy to read."} \\ & P14 & 5 & 4 & 5 & 5 & \textit{"Very slick website, very well organised. Very well designed."} \\
\midrule
\multirow{2}{*}{Navigation} & P06 & 4 & 4 & 4 & 4 & \textit{"I can find my way around."} \\ & P14 & 5 & 4 & 5 & 5 & \textit{"Easy to navigate around, easy to find the information… the search functionality is very good."} \\
\midrule
\multirow{2}{*}{Terminology} & P06 & 5 & 3 & 4 & 4 & \textit{"The terms… I could understand this and if somebody like me is your target audience, then you're OK."} \\ & P14 & 5 & 4 & 4 & 4 & \textit{"Everything is very clear… maybe call \emph{pedagogy} 'teaching, learning and assessment'."} \\
\midrule
\multirow{2}{*}{User Satisfaction} & P06 & 5 & 5 & 5 & 5 & \textit{"[Overall experience] Excellent. Very easy to use."} \\ & P01 & 5 & 4 & 5 & 5 & \textit{"A really good website and something that I will return to… there's so much fantastic content here."} \\
\bottomrule
\end{tabular}
\end{table*}

\subsection{Data Collection}

The  data was collected using the think aloud technique \cite{nielsen2012thinking, Preece2015} followed by a semi-structured interview. Data collection took place in facilitated observational sessions using the concurrent method, which involved live think aloud commentary while performing tasks \cite{Ericsson1993, Alhadreti2017}.
During online or offline sessions (30–90 minutes each), using MS Teams for video recording and automatic transcription, participants completed two representative tasks with relevant subtasks while verbalising their thoughts;  Task 1 (shortened): \textit{find a gender initiative that they could implement}, Task 2 (shortened): \textit{find the most promising gender initiative from the toolkit while assuming that all resources are available for its implementation}. Both tasks were designed to assess how educators locate and interpret content within the toolkit, rather than to evaluate whether they subsequently implement any initiative. The evaluation therefore captures perceived usefulness and usability under conditions of authentic engagement, not evidence of downstream practice change. Claims about actionability in the findings should be understood in this sense: that educators perceived TechMate as sufficiently clear and concrete to support implementation, not that implementation was observed or verified.

Following task completion, interview questions elicited usability and usefulness attributes and included: \textit{How did you find your experience locating and completing the tasks?} (Effectiveness) \textit{Did you trust what you read?} (Trustworthiness) \textit{What could we do better to make the toolkit useful to you?} (Satisfaction) \textit{How could this toolkit help you perform a gender initiative in the real-life scenario?} (Actionability).

Eighteen computing educators (11 women, 7 men) were recruited from eight publicly funded Irish universities through INGENIC (Irish Network for Gender Equality in Computing) and the network's referrals. Participants represented diverse seniority levels and roles within computing departments (outreach leads (1), teaching-focused including senior teaching faculty (10), professors (3) and academic management (4)), ensuring broad representation of the target audience.  
No compensation was provided.

\subsection{Data Analysis}

First, \textbf{meaningful units}, defined as coherent segments capturing single ideas relevant to evaluation aims \cite{Braun2006}, were extracted from the transcripts. 
A meaningful unit could be a single sentence, multiple sentences, or a paragraph that expresses one complete thought relevant to a specific usefulness or usability attribute. For example, \textit{``It's very easy to find things... you see ... and I'm finding things as I would expect''} constitutes one meaningful unit related to navigation, while \textit{``I trusted what I read because it's all seems to be well researched and evidence based, and the evaluation approach is quite important as well because it shows how to assess impact''} would be split into two units (trustworthiness and actionability respectively). Second, coding of the units \cite{Braun2006, Hsieh2005} was applied, deductively assigning meaningful units to predefined usability and usefulness attributes, and inductively coding them for identification of emergent themes. The process resulted in 1027 records: 491 units for usefulness (47.8\%), 378 units for usability (36.8\%), and 158 units for thematic analysis (15.4\%) (using the approach by Braun \& Clarke \cite{Braun2006} with a line-by-line analysis and grouping similar codes into relevant themes and subthemes).

It is worth clarifying the distinction between these two analytical moves. The deductive scoring of meaningful units against predefined usefulness and usability constructs follows the directed content analysis approach \cite{Hsieh2005}: analysis begins with established theoretical categories and applies them to the qualitative data, assessing the direction and intensity of participants' responses within those categories. This is distinct from the inductive thematic analysis \cite{Braun2006}, which started without predefined categories and allowed themes to emerge from participants' broader reflective commentary. The two approaches were applied to separate subsets of the data and serve different purposes: the directed content analysis addresses RQ1 and RQ2 by systematically assessing how participants responded to specific toolkit attributes, while the thematic analysis addresses RQ3 by surfacing unanticipated insights that emerged during authentic task engagement.

Each meaningful unit that was assigned a usefulness or a usability attribute was scored manually by three independent coders on a five-point Likert-type scale \cite{Vagias2006}: \textbf{Very Positive (5)}, \textbf{Positive (4)}, \textbf{Neutral/Mixed (3)}, \textbf{Negative (2)}, \textbf{Very Negative (1)}. 
For example, the statement \textit{``Really good, really good [the overall experience]''} would be scored as Very Positive (5), while \textit{``I thought the layout is OK''} would be scored as Neutral/Mixed (3).
Coders used a shared scoring rubric (Table \ref{tab:sentiment-rubric}) 
and worked independently until completion. Clarifications were resolved in a discussion upon completion. Inter-rater reliability was assessed post-discussion using the intraclass correlation coefficient (ICC) \cite{shrout1979intraclass}, which is appropriate for ordinal data and multiple raters \cite{hallgren2012computing}. The ICC was 0.814 (95\% Confidence Interval [0.752, 0.856]), demonstrating excellent reliability \cite{cicchetti1994guidelines}. The final result for each unit was calculated as the average of scores from three coders and rounded to the nearest full number. This multi-coder approach enhanced credibility while maintaining interpretive depth \cite{Braun2019, MacQueen1998}. An extract from the final scoring sheet is displayed in Table \ref{tab:attribute-scores-quotes-reduced}. 

\begin{table}[t]
\centering
\small
\caption{Sentiment scoring rubric for meaningful units.}
\label{tab:sentiment-rubric}
\resizebox{\columnwidth}{!}{%
\begin{tabular}{clp{3.0cm}p{2.7cm}}
\toprule
\textbf{Score} & \textbf{Label} & \textbf{Anchor description} & \textbf{Example cues} \\
\midrule
5 & Very Positive & Strong, unambiguous endorsement or enthusiasm. & ``I completely trust what I read.'' ``This is brilliant.'' \\[0.6ex]
4 & Positive & Clear approval or favourable evaluation. & ``It's nicely designed.'' ``I'd probably return to this.'' \\[0.6ex]
3 & Neutral/Mixed & Ambivalent, balanced, or informational. & ``It has good material, but I'd need more time to judge.'' \\[0.6ex]
2 & Negative & Clear criticism or dissatisfaction. & ``I'm not sure what I'm supposed to be doing.'' \\[0.6ex]
1 & Very Negative & Strong rejection or frustration. & ``This is pointless; I wouldn't use it at all.'' \\
\bottomrule
\end{tabular}%
}
\end{table}

At least one unit was identified for each attribute per participant. The max number of units identified for an attribute per participant was 24. 
For each participant, an average score was calculated for each attribute, and these participant‑level averages were then used to examine how responses were distributed across attributes. To show the overall average for each attribute across all 18 participants, the mean of these participant‑level scores was computed and rounded to one decimal place.
These scores enabled systematic comparison across attributes and participants.

\section{Results \& Discussion}\label{results}

The following overview of initiative selection provides context for the attribute-level findings reported under RQ1 and RQ2. Participants selected 18 of 26 available initiatives (69\%) across all four toolkit categories, demonstrating broad interest. For Task 1 (implementable initiative), teaching-focused faculty (10 participants) chose pedagogical practices (50\%: class/lab setup, personalised feedback, hybrid programmes) and influence \& support mechanisms (30\%). Academic management/professors (7 participants) selected influence \& support mechanisms and role models (43\%), followed by outreach and teaching practices (28.5\% each). For Task 2 (unlimited resources), 40\% of teaching faculty selected influence \& support mechanisms while 70\% of academic management chose policy-change initiatives.

Overall, influence \& support mechanisms received most selections (12) among this sample of engaged educators, including first-year meetings, buddy systems, and guest lecturers as role models (3 each). Top individual initiatives were gender-balanced outreach events and intelligent class/lab management (4 each), followed by family-friendly schedules, role model lectures, and buddy systems (3 each).

\begin{figure*}[t]
    \centering
    \begin{subfigure}[t]{0.33\textwidth}
        \centering
        \raisebox{2cm}{ %
        \includegraphics[height=3.7cm]{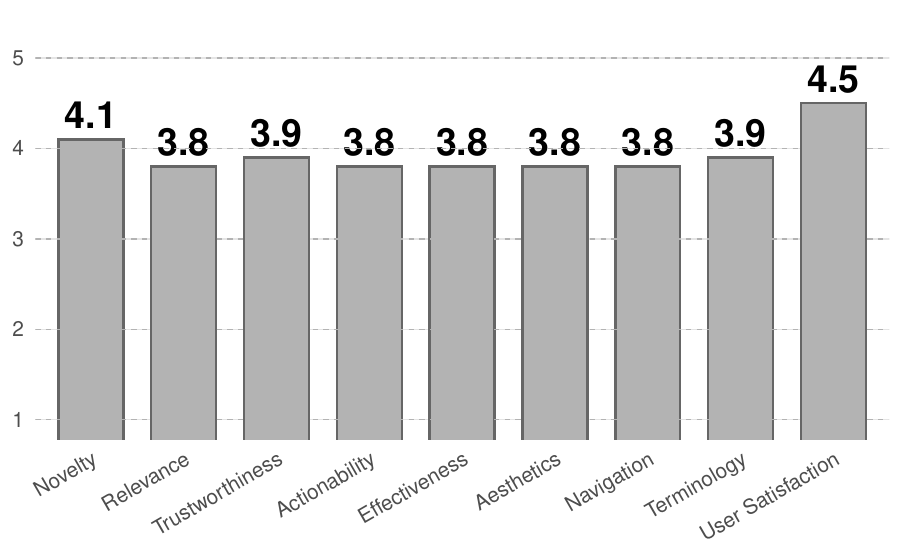}}
        \caption{Average scores across dimensions}
        \label{fig:avg}
    \end{subfigure}
    \hfill
    \begin{subfigure}[t]{0.66\textwidth}
        \centering
        \includegraphics[height=7cm]{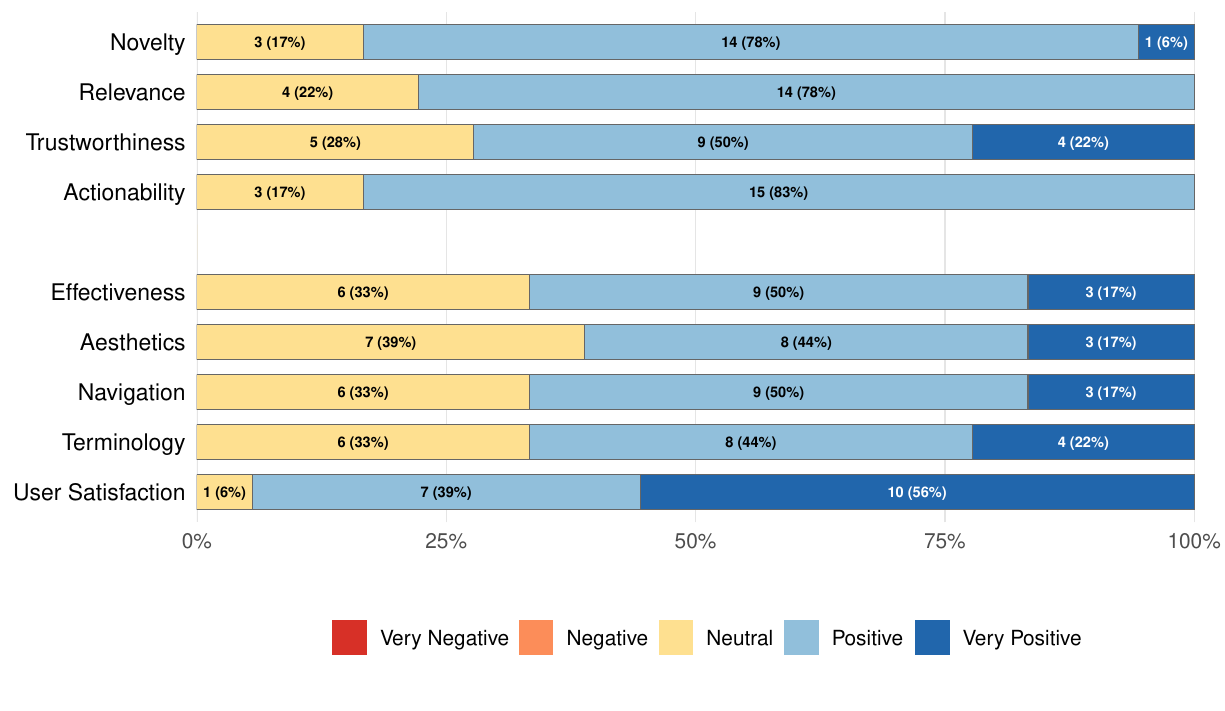}  
        \caption{Distribution of responses}
        \label{fig:dist}
    \end{subfigure}
    \caption{User evaluation results across multiple dimensions.}
    \label{fig:combined}
\end{figure*}

\subsection{Quantitative Scoring}

All attributes were rated positively, with average scores between 3.8 and 4.5 (Figure~\ref{fig:combined} (a)). All attributes received positive or neutral ratings from all participants (Figure \ref{fig:combined} (b)).

\textbf{RQ1: How do educators perceive the usefulness of the toolkit's content?} 

\textbf{Novelty} (120 units) and \textbf{Relevance} (128 units) were rated positively by at least 78\% of participants, indicating educators perceived the toolkit as offering fresh, contextually-aligned ideas. Participants discovered unfamiliar initiatives (e.g., short first-year meetings, course description wording changes) and valued having research-backed strategies consolidated. 

\textbf{Trustworthiness} (62 units) was rated positively by 72\% of participants. Trust stemmed from peer-reviewed research links, known institutional partners, and professional appearance, though some suggested more explicit authorship and institutional endorsements. Trust combined source credibility with transparent evidence trails, which is consistent with established concepts \cite{Wathen2002}.

\textbf{Actionability} (181 units) achieved 83\% positive ratings. Participants valued "ways to implement" and "evaluation approach" sections, noting that concrete steps, templates, and categorisation helped identify feasible initiatives and evidence impact. Strong actionability ratings align with task-technology fit perspectives \cite{Wang1996, Venkatesh2003}. 
Informed by Wang and Strong's consumer-centred data quality framework \cite{Wang1996}, which defines information quality through relevance, timeliness, and interpretability for users' decision making, the results indicate that the toolkit demonstrates strengths across these dimensions. However, based on actionability scores (which are clustered at "positive" rather than "very positive"), it could be enhanced through context-specific decision-making support (e.g., ``if you have limited outreach capacity, start here'').

Task selection patterns further contextualise these findings: realistic choices (Task 1) favoured pedagogical practices within educators' sphere of influence (suggesting that the toolkit is both relevant and interpretable for users in their day-to-day roles), while aspirational choices (Task 2) shifted toward policy-level change (indicating that while users recognise the importance of systemic change, these options may be perceived as less immediately actionable; this is where context specific guidance could support user's ambitions). With 69\% of initiatives selected and strong interest in role models, mentoring, and in-class practices, content proved both evidence-based and personally actionable across institutional roles.

In answer to \textbf{RQ1}, educators perceived the toolkit as genuinely useful: offering novel, role-relevant ideas they could trust and translate into concrete gender initiatives.

\textbf{RQ2: How usable is the toolkit for computing educators?} 

\textbf{User Satisfaction} (43 units, average 4.5) was strongest, with 95\% positive ratings describing the toolkit as ``amazing'' and ``brilliant,'' valuing that ``everything is in one place.''

\textbf{Effectiveness} (92 units) and \textbf{Navigation} (97 units) received 67\% positive ratings; task completion was "straightforward" and "clear." Multiple access points and filtering aided navigation, though some suggested clearer differentiation when drilling into initiatives.

\textbf{Aesthetics} (104 units) and \textbf{Terminology} (42 units) scored 61\% and 66\% positive: design was clean and professional, with requests for larger fonts and occasional tooltips. 
As these were the weakest attributes, a small set of immediate interface refinements have been identified, for example, adding tooltips to explain key terms on first use, improving contrast and fonts, and visually integrating the category menu and filters more clearly. Although accessibility was not formally evaluated in this study, participants’ comments on font size and visual hierarchy highlight its importance. Future work will include systematic accessibility testing and iterative adjustments to ensure that TechMate is usable by a broad range of educators, including those using assistive technologies.

In line with ISO~9241-11 \cite{ISO9241-11} and established usability guidance \cite{Nielsen1993}, analysis of the think aloud data shows that participants could locate initiatives, evaluation approaches, and resources without extensive prompting, indicating adequate cognitive fit for typical tasks \cite{Vessey1991}. At the same time, participants identified a small set of refinements: improved visual hierarchy and labels, modest typography and colour adjustments, tighter integration and signalling of filter options that would further strengthen accessibility, discoverability, and first-time comprehension without requiring major redesign.
In answer to \textbf{RQ2}, the consistently high ratings and think aloud observations indicate that the toolkit provides a highly usable experience for computing educators, who can locate and understand its guidance with minimal effort, while some minor technical improvements are needed to further optimise the interface.

It is important to note that as participants were recruited via a national network for gender equality in computing and referrals, the pool is likely to comprise educators who are already motivated and knowledgeable about gender inclusion. This may partly explain the tight clustering of scores in the positive range: participants may have been predisposed to view a research‑informed gender equality toolkit favourably. As a result, the findings are most directly generalisable to early adopters and champions, and future work should include educators with weaker prior engagement in gender initiatives to test whether the toolkit remains as usable and useful in less favourable conditions.

\subsection{Thematic Analysis}

To note is that the thematic findings extend beyond the toolkit itself, reflecting participants' wider professional experience in computing higher education. This is a productive outcome of \textbf{RQ3}'s design: situating participants in an authentic task, rather than a decontextualised questionnaire, the evaluation surfaced the structural constraints that any implementation-focused tool must navigate.
The inductive analysis of participants’ feedback yielded four themes with relevant subthemes
(Table \ref{tab:compact-themes}).

\begin{table}[h!]
\centering
\caption{Themes and sub-themes with an example unit.}
\footnotesize
\renewcommand{\arraystretch}{0.95}
\begin{tabular}{p{2.3cm}p{5.3cm}}
\toprule
\textbf{Theme} & \textbf{Sub-theme (\#units) and sample unit} \\
\midrule
\textbf{\scshape Structural/ Cultural Barriers}
& \textbf{Challenges with Implementation and/or Evaluation of Initiatives} (68): 
\textit{“I think it's really hard to convince people to drop math as a requirement when they have had to have maths to get into the programme themselves. They can't imagine a world in computing without maths.”} \\
& \textbf{Need/Lack of Women Leaders} (4): 
\textit{“In my department... Staff is very male-dominant. It’s really an issue.”} \\
& \textbf{Women Students’ Experiences} (16): 
\textit{“Isolation is a big issue. If you are in a minority the probability of somebody finishing in advance of
you is increased.”} \\
\midrule
\textbf{\scshape Ideas on How to Improve the Toolkit}
& \textbf{Design Improvements} (16): 
\textit{“It would be useful to indicate what people have already seen and what has changed since they last visited.”} \\
& \textbf{Content Improvements} (22): 
\textit{“As new things like generative AI come out, you would want to see that added here.”} \\
\midrule
\textbf{\scshape Ways to Address Gender Imbalance}
&  (26): 
\textit{“I’d like being able to meet more women...
if we could bring them to schools it would be
really good.”} \\
\midrule
\textbf{\scshape Ways to Improve Teaching and Learning}
& (6): 
\textit{“I meet with students who are having difficulties and we work together to understand problems.”} \\
\bottomrule
\end{tabular}
\vspace{-6pt}
\label{tab:compact-themes}
\end{table}

\textit{Structural/Cultural Barriers}. This theme with three subthemes (Challenges with Implementation and/or Evaluation of Initiatives, Need/Lack of Women Leaders, Women Students' Experiences) is the largest among others and highlights structural and cultural barriers: time and funding constraints, institutional resistance to change, difficulties sustaining and evaluating initiatives, lack of women role models and the isolation and reduced confidence of women in men-dominated classrooms. Participants explicitly noted that having ``everything in one place'' with concrete actions, role‑model ideas, and ready‑to‑use evaluation tools made it easier to tackle issues like isolation, lack of support, and the difficulty of evidencing impact, and several stated they would use the toolkit to justify and implement gender initiatives in their own departments, and some wished for institutional endorsement that would help new users who are not familiar with the toolkit to start using it, as well as the promotion of adoption at other universities.

\textit{Ideas on How to Improve the Toolkit
}covered two main enhancement directions (subthemes) Design and Content Improvements. Participants proposed richer contextualisation (case studies, practitioner stories, short videos) and community features (feedback forms, user-submitted examples) to turn the toolkit into a living repository of ``what works.'' Design suggestions focused on accessibility and trust-building, including clearer authorship, indicators of updated content, and subtle visual refinements.

\textit{Ways to Address Gender Imbalance}. 
Participants reported a wide range of existing and aspirational strategies that they practice or see as promising to do, e.g., outreach, mentoring, ambassador programs, role-model visibility, curriculum redesign, and alternative entry pathways (e.g., removing a math requirement) that are already covered in TechMate and were extensively discussed in the sessions. While the toolkit provides guidance, the additional details and examples such as case studies could highlight practical insights that make these strategies even more valuable.

\textit{Ways to Improve Teaching and Learning.} 
The final theme highlighted student-centred pedagogies, which, as highlighted by the participants, are not gender specific. These included small-group work to learn how to help others, early support for all struggling students, diverse cognitive approaches to programming and framing programming as learnable rather than innate. While effective, these practices are time-intensive, creating tension with existing workloads.

Overall, these themes show that, even though gender‑inclusive practice is limited by structural and cultural barriers, a high‑quality toolkit like TechMate can actively help educators navigate those constraints by offering concrete actions, shared practice‑based evidence, and support for making the labour of inclusion visible and recognised, thereby directly addressing \textbf{RQ3} through insight into educators’ practical needs and contexts. Whether the barriers mentioned by participants could prevent implementation entirely, delay adoption, or diminish over time as educators build evidence of effectiveness remains unknown. Longitudinal research tracking educators over 1-2 academic years could reveal which toolkit actions users progress from browsing to implementation; which structural barriers could be addressed through toolkit-provided evidence and resources versus which require institutional intervention; whether early adopters become institutional champions who reduce barriers for colleagues; and whether sustained use can lead to cultural shifts that normalise gender-inclusive practice.

\section{Conclusion}\label{conclusion}

The evaluation showed that TechMate acts as both a catalyst for local action and a lens on systemic barriers where initiatives meet cultural and institutional resistance.
Results indicate low usability risk for departmental adoption, though effective uptake requires time, recognition, and structures supporting implementation. Several limitations of this work should be acknowledged: the small single-country sample (18 participants, eight Irish universities) may limit generalisability; recruitment through INGENIC means participants were likely already motivated to engage with gender inclusion, which strengthens ecological validity but may have inflated positive evaluations of usefulness and relevance; participants' familiarity with the research context may have introduced social desirability bias, particularly in trustworthiness ratings; the sample can be best seen as representing engaged early adopters rather than a random cross‑section of all computing educators; the evaluation captured initial impressions rather than longitudinal studies; and the third-level focus addresses only one pipeline segment.

Future work includes longitudinal studies to track adoption of the toolkit across national and international contexts; community contribution mechanisms, which allow active collection and publication of case studies; and assessing downstream recruitment and retention impact.

TechMate appears adoptable and empirically supported, with scope for improvement in authorship clarity, visual design, filter discoverability, and community features. As an open-access resource freely available to educators across institutional contexts, it illustrates both the potential and the limits of technology-based tools for social good: the technology can lower the barrier to evidence-informed practice, but institutional culture, workload, and recognition structures shape whether that potential is realised. Addressing these conditions is as much a governance and policy challenge as a design one.

\begin{acks}

Generative AI tools (ChatGPT, Claude, Sana) were used for wording suggestions, LaTeX/BibTeX formatting, and minor code edits; all research design, analysis, results, and final text are the work of the human authors. This work was supported by the HEA in Ireland, Huawei Ireland, and Technological University Dublin.
\end{acks}

\bibliographystyle{ACM-Reference-Format}
\bibliography{sample-base}

\end{document}